\documentclass[12pt]{article}
\usepackage{pdproc} 
\usepackage{epsfig} 

\begin{document}

\renewcommand{\thefootnote}{\alph{footnote}}
  
\title{DIFFUSE NEUTRINO FLUX}

\author{J\"URGEN BRUNNER}

\address{ CPPM, Aix-Marseille Universit\'{e}, CNRS/IN2P3, Marseille, France;\\
on leave of absence at DESY, Platanenallee 6, Zeuthen, Germany}

\abstract{The search for a diffuse neutrino flux component from astrophysical
sources complements searches for point sources. In 2010 and 2011 many
new results have been published on this subject. 
Realistic models can now be tested with these new measurements 
from the leading neutrino telescope experiments. 
An overview over these recent results is given.}

\normalsize\baselineskip=15pt

\section{Introduction}

Neutrinos are produced in a variety of sources throughout the universe. The most
prominent natural neutrino point source is our Sun. By exploring the angular correlation
between the neutrino arrival direction and the position of the Sun, the solar origin 
of these neutrinos could be undoubtedly proven. 
However apart from our Sun no other permanent neutrino sources could be identified so far.
An alternative is the search for a diffuse neutrino flux. Individual sources cannot
be distinguished but the spectral shape of their integral contribution might be used
to distinguish them from possible background. Such a diffuse neutrino flux might exist at
very different energies. 

A large contribution is predicted from neutrinos which decoupled
from thermal equilibrium at a temperature of the universe of about 1~MeV. Due to the
expansion of the universe these neutrinos (equipartitioned over all flavours) are expected
to have today a black body radiation spectrum with a mean energy in the meV region.
From measurements of neutrino oscillations it is known that the two heavier neutrinos 
have masses of $m_2>9$~meV$/c^2$ and $m_3>50$~meV$/c^2$. Therefore they must be
non-relativistic today. So far no experimental technique is known to detect them.

Another diffuse neutrino flux must exist when summing the neutrino yield of all stellar fusion 
processes in the history of the universe. These neutrinos are expected to have spectra
similar to solar neutrinos with energies mainly in the region of 0.1~MeV to 10~MeV. 
At slightly higher energies (up to 50~MeV) one should find a diffuse flux from neutrinos 
which had been produced during neutron star formation at core collapse Supernovae explosions 
throughout the history of the universe. A search for such a  neutrino flux has been done in the
SuperKamiokande experiment~\cite{SuperK-diff}. No signal is observed and a flux limit could be
set. This is discussed in more detail elsewhere in these proceedings~\cite{SuperK}.

In the following we concentrate on the diffuse flux at energies $E_\nu > 1$~TeV. Only the
most violent processes in the universe such as Active Galactic Nuclei (AGNs) or Gamma Ray
Bursts (GRBs) can contribute in this energy range. Its observation requires significantly
larger detectors compared to those found in underground sites. In the following, results 
from large scale Neutrino Telescopes are presented which use natural transparent media 
such as deep sea water or Antarctic ice. A detailed introduction into the 
concept and physics of these devices is found elsewhere in these proceedings~\cite{NT}. 
Here results from the Baikal~\cite{Baikal-diff-casc}, AMANDA~\cite{Amanda-diff-mu}, 
ANTARES~\cite{Antares} and IceCube~\cite{IceCube} Neutrino Telescopes will
be discussed. 

\section{The cascade and the muon channel}
\label{signature}
Neutrino interactions in the vicinity of the detector lead to two distinct signatures 
which are exploited in independent analyses. The charged current interaction of 
muon neutrinos
\begin{equation}
\nu_\mu(\bar\nu_\mu)+N\rightarrow \mu^\mp+X
\end{equation}
and the charged current interaction of tau neutrinos with a subsequent muonic decay
\begin{eqnarray}
&\nu_\tau(\bar\nu_\tau)+N\rightarrow &\tau^\mp+X \\
& &\tau^\mp\rightarrow\mu^\mp+\bar\nu_\mu(\nu_\mu)+\nu_\tau(\bar\nu_\tau)
\end{eqnarray}
result both in a long muon track. At TeV and even more at PeV energies these muons have 
a range of several km in water or ice, largely exceeding the size of the detectors.
The neutrino interaction vertex and the accompanying hadronic shower are outside the 
fiducial volume most of the time. The track signature yields a good angular resolution.
This is in principle not needed for a search of a diffuse flux
but it ensures a clean separation of an upward going neutrino signal from 
the background of downward going atmospheric muons. The muon energy at the detector 
is deduced from the energy loss dE/dx which is related to the ``brightness" of the track in the
detector. As the energy loss is mostly stochastic at TeV/PeV energies, the muon energy can 
only be estimated with a precision of about a factor two. 
The calculation of the related neutrino energy depends on the assumed neutrino flux
because the muon track has to be extrapolated upstream the detector to the neutrino vertex.

All other neutrino interactions as neutral current reactions
\begin{equation}
\nu_x(\bar\nu_x)+N\rightarrow \nu_x(\bar\nu_x)+X
\end{equation}
electron neutrino charged current interactions
\begin{equation}
\nu_e(\bar\nu_e)+N\rightarrow e^\mp+X
\end{equation}
and charged current interactions of tau neutrinos with a subsequent non-muonic decay
\begin{eqnarray}
&\nu_\tau(\bar\nu_\tau)+N\rightarrow &\tau^\mp+X \\
& &\tau^\mp\rightarrow\nu_\tau(\bar\nu_\tau)+X
\end{eqnarray}
lead to a so-called cascade signature with the exception of tau neutrino charged current
interactions at energies above 10~PeV which would yield two distinct cascades, one at the
neutrino interaction vertex and a second one at the tau decay point. 
This special case is not considered in the following.

As a result of these neutrino interactions one obtains a hadronic and, 
depending on the channel, also an electromagnetic shower. 
They are very narrow and have a longitudinal extension of at most 
a few tens of meters. Due to the large spacing of adjacent detector elements in the 
coarsely equipped  neutrino telescopes it is impossible to distinguish electromagnetic from
hadronic showers. The observable signature in the detector is in all cases 
an isolated ``cascade". Due to its small extension the angular resolution is much worse here
than for the track signature. This compromises the up/down separation. The main background for
the cascade search comes from bright electromagnetic showers (e.g. due to bremsstrahlung) which
accompany downward going atmospheric muons. The energy resolution is instead much better than for the
track channel. All particles but the escaping neutrinos are seen in the detector and the
brightness of the events correlates directly to the cascade energy which in turn is closely
related to the neutrino energy. When using containment conditions for the neutrino vertex
an energy resolution of 20\% is feasible in future studies in this channel.

Whereas the effective volume for the cascade channel is close to the equipped
volume of the detector, it is significantly larger for the muon channel
due to the very long muon range. This results in a better sensitivity of the
track channel for searches. 
The full potential of the cascade channel is reached at PeV energies where 
the background from electromagnetic activities which accompany downward going 
muon tracks is less dominant.

\section{Atmospheric Neutrinos}
 
The major background for observing a TeV-scale diffuse neutrino flux consists
in atmospheric neutrinos and downward going atmospheric muons. 
Atmospheric neutrinos are produced over a large energy range in interactions 
of primary Cosmic Ray particles (mainly protons) with nuclei in the Earth atmosphere.
The resulting hadronic showers contain also short-lived particles like pions and kaons.
The main sources of conventional atmospheric neutrinos are their decays
\begin{eqnarray}
\label{k-pi}
&\pi^\pm &\rightarrow \mu^\pm+\nu_\mu(\bar\nu_\mu)  \\
&K^\pm &\rightarrow \mu^\pm+\nu_\mu(\bar\nu_\mu)  \\
&K^\pm &\rightarrow e^\pm+\nu_e(\bar\nu_e)+\pi^0.
\end{eqnarray}
The subsequent decay of the muon contributes only marginally to the multi-GeV neutrino flux
as most of these muons reach the ground and they are stopped before decaying. 
The primary Cosmic Rays have a non-thermal $E^{-\gamma}$ spectrum with $\gamma\approx 2.7$.
As pions and kaons propagate a certain distance through the atmosphere and lose thereby 
energy before decaying, the resulting neutrino spectrum is softer 
with $\gamma\approx 3.7$.

The most recent published measurements of the high energy part of the atmospheric neutrino spectrum 
comes from the IceCube collaboration~\cite{IC40-diff-mu,IC40-atm-nu}. 
One year of data (2008-2009) has been analysed with almost half of the final 
detector lines installed (40 out of 86). As much as 18,000 upward going 
$\nu_\mu$ candidate events have been selected for this analysis. 
So far only the track signature has been exploited to measure the 
atmospheric neutrino flux above 100~GeV. 
This has several reasons. First $\nu_\mu$ are more copiously
produced than other flavours (see Equation~8-10) and neutral current interactions have a
lower cross section than charged current reactions. Further the effective volume is larger
for the muon track signal than for the cascade channel and finally the isolation  
of a clean upward going event
sample is more difficult in the cascade channel, as discussed above.

Two methods have been used to extract the neutrino spectrum from the data, 
forward folding~\cite{IC40-diff-mu} and regularized unfolding~\cite{IC40-atm-nu}. Both
methods give comparable results and they are also consistent with different conventional
atmospheric neutrino flux calculations~\cite{bartol,honda}. 
A distinction between these different flux predictions is not possible 
within the precision of the measurement. The highest energetic events can be attributed to
neutrinos with energies of more than 100~TeV, which are thereby the highest energetic
neutrino interactions ever detected. 

For neutrino energies above 10~TeV the decay of mesons which contain heavy quarks (c,b) 
starts to contribute to the atmospheric neutrino flux. As these mesons have typical 
decay lengths of only few mm, they do not lose energy before they decay and the resulting 
so-called
prompt atmospheric neutrino spectrum follows closely the original Cosmic Ray spectrum, {\it i.e.} 
$\gamma\approx 2.7$. This should be seen as a hardening of the measured spectrum and has
been searched for in one of the mentioned IceCube analyses~\cite{IC40-diff-mu}. 
The predictions for this prompt neutrino flux vary by up to a 
factor~ten~\cite{prompt-1,prompt-2,prompt-Bugaev-RQPM,prompt-Naumov-RQPM,prompt-Costa-RQPM}. 
The largest contribution is obtained in the frame of the 
Recombination Quark Parton Model~\cite{prompt-Bugaev-RQPM,prompt-Naumov-RQPM,prompt-Costa-RQPM} 
(RQPM), a non-perturbative QCD approach. 
As no hardening of the neutrino spectrum is observed, 
the RQPM model from~\cite{prompt-Bugaev-RQPM} can be excluded on a level of 3$\sigma$ 
in~\cite{IC40-diff-mu}. Further the model from~\cite{prompt-2} contains free parameters.
When choosing these parameters to maximize the prompt neutrino flux, this model can be
excluded with 2$\sigma$. 
The model from~\cite{prompt-1}, which predicts the lowest prompt flux,
can instead not be constrained by the current data.

\section{Search for a diffuse neutrino flux from astrophysical sources}

An upper bound for a diffuse neutrino flux from astrophysical sources has been
derived by Waxman and Bahcall (W\&B)~\cite{WB-limit}. Here it is assumed that the
extragalactic Cosmic Ray spectrum for $E > 10^{18}$~eV is produced
in sources where protons are magnetically 
confined to undergo efficiently the photoproduction reaction 
\begin{equation}
p+\gamma\rightarrow\Delta^+\rightarrow n+\pi^+.
\end{equation} 
The pions decay according to Equation~8 and produce neutrinos, whereas the
neutrons escape from the acceleration site, decay and produce the
observed high energetic Cosmic Ray spectrum. Therefore the predicted neutrino
flux is closely related to the observed Cosmic Ray flux above $E > 10^{18}$~eV
and should be proportional to $E^{-2}$ over several orders of magnitude.
The resulting upper bound (corrected for neutrino oscillations 
during propagation from the source to Earth) is shown on Figure~\ref{limit}. 
Other models try to circumvent the
constraints of~\cite{WB-limit} and predict higher neutrino fluxes (see below). 
By now the sensitivity of the experiments is high enough to test some 
of these models.

\subsection{The cascade signature}

A search for cascade signatures from a diffuse neutrino flux has been performed
by Baikal~\cite{Baikal-diff-casc}, 
AMANDA~\cite{Amanda-diff-casc,Amanda-diff-casc-UHE} and
IceCube~\cite{IC22-diff-casc}. The details of theses analyses vary. But one 
common feature is the major discriminating power of the event brightness,
expressed in terms of hit counts or amplitude, which
is used to distinguish the signal from background. 
For all these analyses the main
background comes from electromagnetic showers, associated to downward 
going muons.
None of the analyses finds an excess of events over the background
expectation. This allows to derive limits for an $E^{-2}$ signal flux. 
\begin{table}[htpb]
\begin{center}
\begin{tabular}{||l|l|r|r|l|l|l||} \hline
Experiment & Data taking & \multicolumn{2}{c|}{Energy range} 
& \multicolumn{2}{c|}{$E^2\Phi$} \\ \hline
Baikal \cite{Baikal-diff-casc}   & 1998 - 2002   & 20 TeV & 20 PeV    & $2.9\cdot 10^{-7}$ & $9.7\cdot 10^{-8}$  \\
AMANDA \cite{Amanda-diff-casc}  & 2000 - 2004   & 40 TeV & 9 PeV     & $5.0\cdot 10^{-7}$ & $1.7\cdot 10^{-7}$\\
AMANDA \cite{Amanda-diff-casc-UHE} & 2000 - 2002   & 200 TeV & 1000 PeV & $2.4\cdot 10^{-7}$ & $8.0\cdot 10^{-8}$\\
IceCube \cite{IC22-diff-casc} & 2007 - 2008   & 24 TeV & 6.6 PeV   & $3.6\cdot 10^{-7}$ & $1.2\cdot 10^{-7}$ \\ \hline
\end{tabular}
\end{center}
\caption{Comparison of the 90\% c.l. diffuse flux limits for different 
experiments in the cascade channel. The rightmost column indicates the single-flavour 
limit which is derived from the all-flavour limit by applying a factor 1/3.
The fluxes are given in (GeV cm$^{-2}$s$^{-1}$sr$^{-1}$). }
\label{diff-casc}
\end{table}
Table~\ref{diff-casc} compares the diffuse flux limits from these analyses.
The second column shows the corresponding data taking period, 
the third and fourth
columns the central energy range in which 90\% of 
the signal events would be found. AMANDA has performed a special analysis
searching for downward going ultra high energetic (UHE)
events~\cite{Amanda-diff-casc-UHE}. This yields the best current limit for
energies above 200~TeV.
As explained in section~2 all
flavours contribute to the cascade channel. 
As neutral current reactions are flavour-blind, there
is an exact equipartition between all flavours in this channel, 
whereas for charged current
reactions the main (but not exclusive) contribution comes from $\nu_e$.
Therefore limits in column~5 are given as ``all flavour" limits, assuming that
the flux at the detector is composed from all three neutrino flavours and both
neutrino helicities in equal parts. The combination of 
Baikal~\cite{Baikal-diff-casc} and the AMANDA UHE~\cite{Amanda-diff-casc-UHE}
results give the best limits over the whole energy range. To make these results
compatible to the limits derived in the $\nu_\mu$ track channel, they are 
converted to single flavour limits by simply dividing by a factor 3. 
The corresponding single flavour limits are given in the last column 
of Table~\ref{diff-casc}.  

\subsection{The muon track signature}

AMANDA~\cite{Amanda-diff-mu}, ANTARES~\cite{ant-diff} and 
IceCube~\cite{IC40-diff-mu} have performed searches for a diffuse flux in the
muon channel. Track reconstruction is used to select a clean sample of upward
going neutrino candidates. The energy is estimated by evaluating the
``brightness per track length" which correlates with the muon energy loss and
therefore the muon energy in the detector. Different methods are applied in the
three analyses. 
A particularly simple one is used in ANTARES. Here the energy is estimated
from the average hit multiplicity on those optical modules which are used 
in the track fit. A cut in this variable is used to select the final event
sample~\cite{ant-diff}. 9~events remain to be compared to expected 8.7 from
conventional atmospheric neutrinos~\cite{bartol} and 2 from the most optimistic
prompt neutrino flux in the frame of the RQPM~\cite{prompt-Costa-RQPM}.
Unlike in the IceCube analysis~\cite{IC40-diff-mu} prompt models cannot be constrained
but limits for astrophysical $E^{-2}$ fluxes are derived.

\begin{table}[htpb]
\begin{center}
\begin{tabular}{||l|l|r|r|l||} \hline
Experiment & Data taking & \multicolumn{2}{c|}{Energy range} 
& $E^2\Phi $ \\ \hline
AMANDA \cite{Amanda-diff-mu}  & 2000 - 2003   & 16 TeV & 2.5 PeV & $7.4\cdot 10^{-8}$ \\
ANTARES \cite{ant-diff} & 2007 - 2009   & 20 TeV & 2.5 PeV & $5.3\cdot 10^{-8}$ \\
IceCube \cite{IC40-diff-mu} & 2008 - 2009   & 35 TeV & 7 PeV   & $8.9\cdot 10^{-9}$ \\ \hline
\end{tabular}
\end{center}
\caption{Comparison of the 90\% c.l. diffuse flux limits 
for different experiments in the $\nu_\mu$ channel. The fluxes are given in
(GeV cm$^{-2}$s$^{-1}$sr$^{-1}$).}
\label{diff-mu}
\end{table}
Table~\ref{diff-mu} compares the diffuse flux limits  
from the three analyses. The data taking periods are indicated in the second column.
It is noteworthy that the ANTARES~\cite{ant-diff} and IceCube~\cite{IC40-diff-mu} results have
been published with a very short delay after data taking. The energy range indicates (as for
Table~\ref{diff-casc}) the central range in which 90\% of the signal events are expected.
The limits in the last column are for $\nu_\mu$, they are single flavour limits
directly comparable to the limits in the last column of Table~\ref{diff-casc}. It can be seen
that the limits obtained in the muon channel are significantly better than the cascade limits.
The larger effective volume and a better discrimination against background from downward 
going muons are responsible for this effect.

\begin{figure}[htpb]
\centering
\epsfig{figure=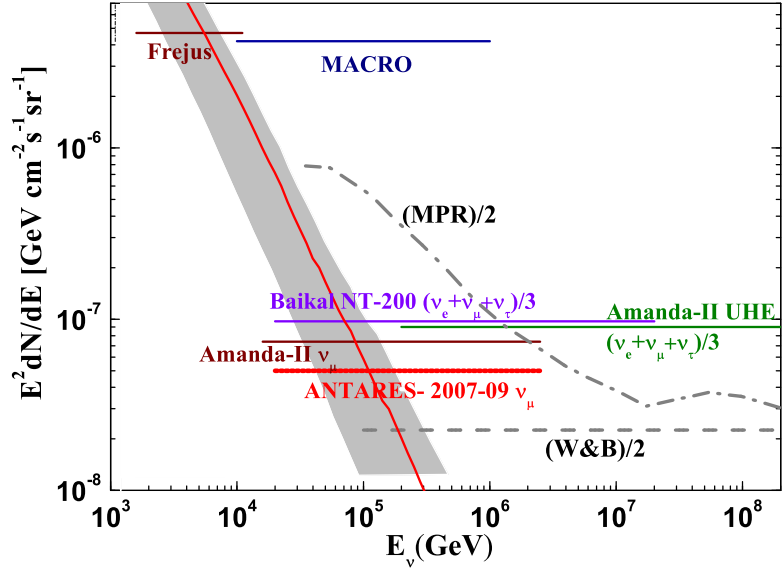,width=12.0cm}
\caption{
A comparison of published 90\% c.l. upper limits for $E^{-2}$ diffuse neutrino
fluxes as of early 2011. For details, see text.} 
\label{limit}
\end{figure}

Figure~\ref{limit} shows a comparison of recent published limits for 
$E^{-2}$ diffuse neutrino fluxes as of early 2011. 
The Frejus~\cite{frejus}, MACRO~\cite{macro}, 
AMANDA-II-$\nu_\mu$~\cite{Amanda-diff-mu} and ANTARES~\cite{ant-diff} limits
have been obtained in the $\nu_\mu$ channel. The Baikal~\cite{Baikal-diff-casc}
and AMANDA-II UHE~\cite{Amanda-diff-casc-UHE} limits 
are from measurements in the cascade
channel. For reference, the W\&B~\cite{WB-limit} and the MPR~\cite{MPR} upper
bounds for transparent sources are also shown. They are divided by two, to take into
account neutrino oscillations. The grey band represents the expected variation of
the atmospheric  flux: the minimum is the Bartol flux~\cite{bartol} 
from the vertical direction; the maximum the Bartol$+$RQPM flux~\cite{prompt-Costa-RQPM} 
from the horizontal direction. The central line
is averaged over all directions.  The figure has been taken from \cite{ant-diff}.

The ANTARES limit constrains various models. The MPR model~\cite{MPR}, shown on 
Figure~\ref{limit} can be excluded with 90\% c.l.. 
Similarly the models~\cite{protheroe,semikoz} are excluded with the same confidence level.

The most stringent limit is currently provided by the preliminary analysis from
IceCube~\cite{IC40-diff-mu} which is not included in Figure~\ref{limit}. It is the 
first analysis which is sensitive enough to probe the region below the W\&B~\cite{WB-limit}
upper bound. The models which are excluded by the ANTARES analysis on 90\% c.l. are here
excluded on a 5$\sigma$ level, as well as some other flux 
predictions~\cite{mannheim,stecker,becker}.

\section{Conclusion}

During the last year a wealth of new results have been published
on the search for diffuse neutrino fluxes from astrophysical sources. These searches
complement point source searches. So far no excess of events beyond the expected rate of
atmospheric neutrino events has been found. The sensitivity of the analyses is already high
enough to constrain or exclude various models. For the first time fluxes below the
W\&B upper bound~\cite{WB-limit} are tested. Fast publication has become a standard
which illustrates a good understanding of the used detectors and media.

\end{document}